\documentclass{article}
\usepackage[utf8]{inputenc}
\usepackage{graphicx}
\usepackage{amssymb}
\usepackage{url,lineno,microtype}
\usepackage{booktabs}
\usepackage{hyperref}
\usepackage{subcaption}
\usepackage{authblk}

\title{Exploring Post COVID-19 Outbreak Intradaily Mobility Pattern Change in College Students: a GPS-focused Smartphone Sensing Study}
\author[1]{Congyu Wu\footnote{Correspondence: congyu.wu@austin.utexas.edu}}
\author[2]{Hagen Fritz}
\author[3]{Cameron Craddock}
\author[2]{Kerry Kinney}
\author[4]{Darla Castelli}
\author[1]{David M. Schnyer}
\affil[1]{Department of Psychology, University of Texas at Austin}
\affil[2]{Department of Civil, Environmental, and Architectural Engineering, University of Texas at Austin}
\affil[3]{Department of Diagnostic Medicine, University of Texas at Austin}
\affil[4]{Department of Kinesiology and Health Education, University of Texas at Austin}

\date{April 2021}

\begin{document}

\maketitle

\begin{abstract}

With the outbreak of the COVID-19 pandemic in 2020, most colleges and universities move to restrict campus activities, reduce indoor gatherings and move instruction online. These changes required that students adapt and alter their daily routines accordingly. To investigate patterns associated with these behavioral changes, we collected smartphone sensing data using the Beiwe platform from two groups of undergraduate students at a major North American university, one from January to March of 2020 (74 participants), the other from May to August (52 participants), to observe the differences in students' daily life patterns before and after the start of the pandemic. In this paper, we focus on the mobility patterns evidenced by GPS signal tracking from the students' smartphones and report findings using several analytical methods including principal component analysis, circadian rhythm analysis, and predictive modeling of perceived sadness levels using mobility-based digital metrics. Our findings suggest that compared to the pre-COVID group, students in the mid-COVID group generally (1) registered a greater amount of midday movement than movement in the morning (8-10am) and in the evening (7-9pm), as opposed to the other way around; (2) exhibited significantly less intradaily variability in their daily movement, and (3) had a significant lower correlation between their mobility patterns and negative mood. 

\end{abstract}

\section{Introduction}

Since the first cases of COVID-19 were confirmed in the United States in January 2020, organizations and individuals scrambled to come up with counter-measures to curb the spread of the virus. State and municipal authorities declared emergency or disaster status, issued stay-at-home orders, and canceled events. Colleges and universities around the country also started implementing decisive measures such as closing campuses and shifting instruction online. One direct consequence is the altered mobility patterns of nearly all people. The way in which people's daily routine changed was unprecedented. College students are a group of people especially influenced by the changes. 

One way to assess mobility patterns is through the use of questionnaires or momentary assessment prompts \cite{depp2019gps}. These methods have their limitations which could be overcome through the use of smartphone sensors, which provide an objective way to measure daily mobility behavior. College students would be an ideal population to examine this approach since they are heavy users of smartphones -- carrying their smartphones with them everywhere they go. A number of smartphone sensing studies have been conducted to monitor health and behavior of college students \cite{wang2014studentlife, 10.1093/gigascience/giab044}. Among smartphone embedded sensors, GPS data tracked over a continuous period of time is found to reflect fluctuations in daily movement patterns and used to detect other related behavior and health issues. We believe smartphone sensing is an appropriate approach to study the mobility change in college students due to the COVID-19 pandemic and the related societal response. 

We collected smartphone sensing data from two groups undergraduate participants from the University of Texas at Austin to investigate mobility pattern change. One group was assessed from January to March, a period that resides mostly before the pandemic was officially declared. The second group was collected from May to August of 2020, during which the pandemic was in full force. To examine mobility and changes in mobility, we applied multiple analytic approaches -- principal component analysis, circadian rhythm analysis, and digital phenotyping -- to quantify the intradaily patterns of participants' movement. Existing studies tend to quantify mobility as a singular construct (e.g., the amount of movement) and the corresponding findings simply suggest ``after the COVID outbreak that people traveled way less everyday". We wanted to utilize more sophisticated GPS data processing techniques to tease out finer patterns that would reflect changes in mobility patterns across the entire 24-hour period, rather than overall movement comparisons. The focus on college students, the use of mobile sensing to study mobility, together with the deep dive into the intradaily patterns are how this paper is unique in the current literature on COVID-19 related behavior change.

Our findings suggest that compared to the pre-outbreak group, students in the post-outbreak group (1) registered greater movements midday rather than in the morning (8-10am) or in the evening (7-9pm); (2) exhibited significantly less intradaily variability in their daily movement, and (3) had a significantly lower correlation between their mobility patterns and their depression symptoms. The first two findings portray a image of a freer and less organized day experienced by the post outbreak group. The third takeaway suggests that mental health prediction using personal sensing features is greater when participants' daily life has more normalcy, that is before the pandemic started.

\section{COVID-19 and Mobility Change}

Researchers have sought evidence of COVID-19-induced mobility change from commercial mobile location data providers. Smartphone apps make location requests sporadically, resulting in a record of GPS locations registered by the smartphone. Commercial location data companies have collected anonymized location pings from an extremely large number of smartphones in the US and globally. An outstanding advantage of such mobile location data is its extensive geographic coverage, allowing researchers to aggregate the data based on specific regions of choice (e.g., city or county) and derive region-specific, population-level insights. The limitation is also significant. Because smartphone location pings are sporadic ``snapshots", the resulting data does not form a continuous portrait of the user's daily mobility pattern and thus cannot be used to understand individual-level behavior. We found several studies using data of this kind to evaluate population mobility change specific to geographic or administrative divisions. Using commercial mobile location data collected shortly after the COVID-19 outbreak, Warren et al. \cite{warren2020mobility} showed a sharp decrease in population mobility in multiple major cities around the world. Gao et al. \cite{gao2020mapping} mapped county-level mobility change in the United States and also found mobility decreases in the vast majority of counties in the early months of rapid COVID-19 spread. Engle et al. \cite{engle2020staying} discovered significant correlations between county-level mobility change and county-level infection rate and other socioeconomic indicators such as age and political affiliation. Couture et al. \cite{couture2021jue} investigated population movement between counties and states in the United States as well as visits to different types of commercial venues. 

While the commercial data approach uncovers population level mobility patterns, a personal sensing approach proves useful to monitor individual participants' daily movement more closely, together with other behavioral aspects such as physical activity and sleep. Sun et al. \cite{sun2020using} collected smartphone sensing (GPS, Bluetooth, phone usage; no accelerometer) and Fitbit data from a large, multi-national sample (1062 participants) in Europe from early 2019 to mid 2020. They found significant decreases in daily distance traveled and the number of surrounding devices detected and increases in phone usage, sleeping time, and home stay after the COVID-19 outbreak. Sa{\~n}udo et al. \cite{sanudo2020objectively} collected smartphone sensing data (accelerometer and phone usage; no GPS) from 20 college students during two periods, once before and once during the COVID-19 lockdown and arrived at similar findings that are reduced physical activity, increased smartphone use, and longer sleeping hours. Most other studies investigating COVID-related behavioral changes in colleges students focused on constructs of physical activity and used questionnaire-based methods \cite{lopez2020impact}. To the best of our knowledge, no existing studies have used objectively measured location data to look into the intradaily patterns of college students as they experience the COVID-19 pandemic. 

\section{Mining GPS Data from Personal Sensing}

Analytical methods for processing GPS data captured by smartphones and other smart wearable devices largely fall in two categories. Both approaches seek to convert the raw GPS trace captured over a period of time (e.g., a day) to a vector of feature values to feed as input to subsequent analyses such as predictive modeling or clustering. The first category is a raw data approach, which directly makes use of the raw GPS data collected and bulk-calculate descriptors or statistics using established algorithms or feature representation methods. This approach aims to preserve information contained in the raw GPS data and requires minimal researcher input on how to manipulate the data. One example of this approach is the vector space representation of GPS locations \cite{eagle2009eigenbehaviors}, which creates a location label (e.g., home, work, etc.) for every predetermined interval (e.g., 30 minutes) during the length of the GPS trace, thus forming a vector of labels characterizing the sequence of place types. The vector space method has also been used to represent other types of mobile sensor signals such as Bluetooth \cite{wu2018vector}. Another example of this approach is training autoencoders using a displacement vector created by differencing the original GPS coordinate series \cite{mehrotra2018using}. A key commonality of the methods that belong in this category is the division of a GPS trace into a series of sub-traces, thus preserving the characteristics of the original trace, and use simple measures of the sub-traces themselves as features without aggregation. 

The second category is a feature engineering approach, which requires direct input from the researcher to devise metrics. Researchers have proposed various GPS features such as location variance, maximum distance covered, and percentage of time spent at home \cite{canzian2015trajectories, saeb2015mobile}, which can be further linked with behavioral and health outcomes and become \textit{digital health phenotypes}. Because they are created with specific purposes to quantify specific constructs, these features are easily interpretable.

\begin{figure}
    \includegraphics[width = \textwidth]{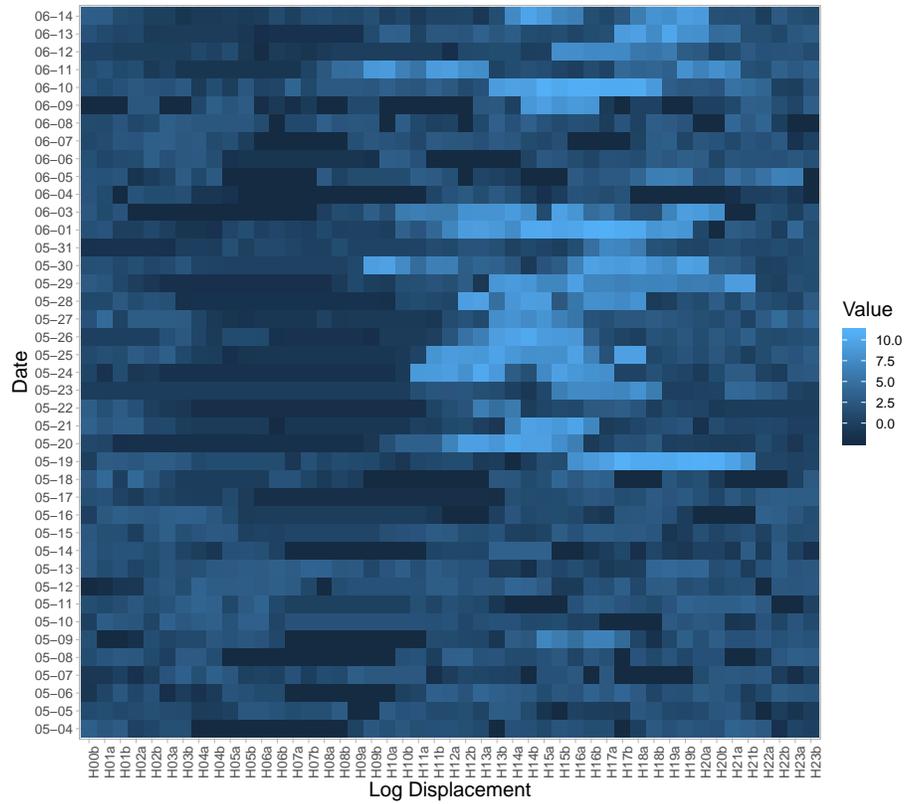}
    \caption{Daily Displacement Profile extracted from one participant's GPS data. Labels on the horizontal axis indicate the half-hour bins: for example, H06a indicates the 6-6:30am and H20b indicates 8:30-9pm. Cell color indicates the natural logarithm of the displacement detected by the smartphone GPS sensor during the current half-hour bin compared to the previous one. Brighter blue means greater displacement.}   
    \label{fig:ddp}
\end{figure}

\section{Data}

We used the Beiwe research platform to collect smartphone sensing and real-time survey data from two groups of undergraduate students at the University of Texas at Austin (UT), over two non-overlapping periods in 2020. Student participants underwent an initial screening before being consented into either phase of the study. Enrollment for Phase 1 started in early January prior to the spring semester. The study period lasted from mid January to the end of March, corresponding to the period of time when the first cases of COVID-19 were confirmed in the US and no nation-wide counter measures had been taken. This study phase included 74 participants. Enrollment interviews for Phase 2 were conducted over a period of two weeks with full enrollment completed by May 1st. The second phase of the study concluded when participants scheduled a virtual meeting with a study coordinator in late August to early September for an exit interview and to coordinate shipping study materials back to UT. Phase 2 consisted of 52 participants during which the pandemic and the related orders and mandates were in full effect.

Smartphone sensing data we collected include GPS, accelerometer, and phone usage data from the participants' primary smartphones and real-time survey data includes participants' responses to daily activity, mood, and sleep questions. Specifically, the GPS data contain timestamped coordinates (longitude and latitude). The GPS sensor was configured to scan for one minute every 10-minute break, subject to hardware constraints such as phone power-off or user deactivating GPS. In total, we collected 6442 days of GPS data across all 126 participants striding the two groups.

\section{Methods}\label{sec:method}

We implemented three different methods to analyze smartphone GPS data collected from our participants in the pre- and post-outbreak groups: principal component analysis, circadian rhythm analysis, and digital phenotyping. The first two belong in the raw data approach whereas the third method involves feature engineering and predictive modeling. For the first two methods, we preprocessed the GPS data by constructing a Daily Displacement Profile for each day of each participant during which GPS data was collected. For each participant's each day's GPS data, we placed the entire GPS trace into 48 half-hour bins and calculate an average coordinate within each bin. Then we calculated the haversine distance between two adjacent bins and regard the vector of subsequent distance values as constituting a Daily Displacement Profile (DDP). A DDP consists of 47 displacement values because of the distance differencing. If during a half-hour bin, no GPS data was observed, then we carried over the coordinate from the most recent available location. Figure \ref{fig:ddp} shows the DDP of all the days collected from one example participant. Overall, we observe increased displacement values on some days during the day compared to in the early morning.

\begin{figure}[t]
    \includegraphics[width = \columnwidth]{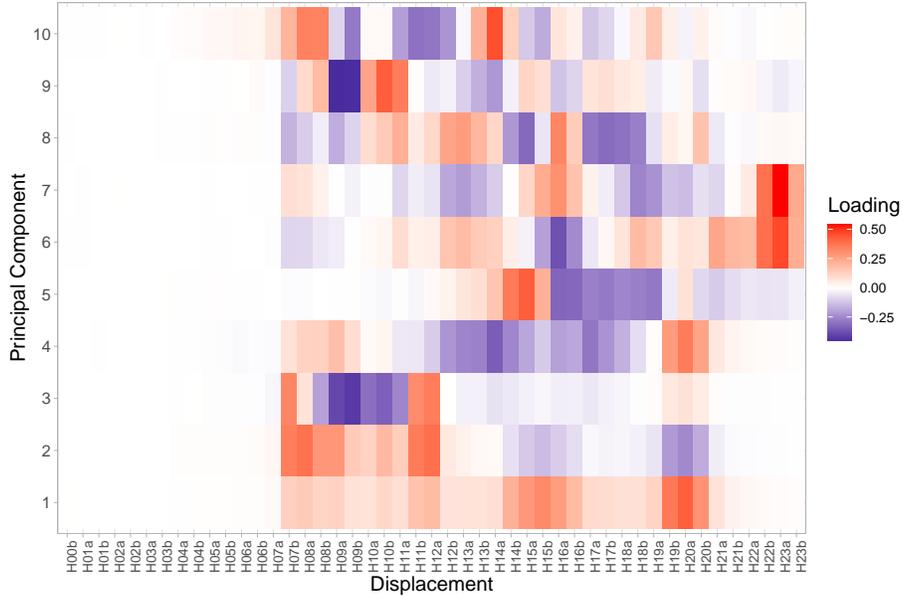}
    \caption{Loadings of the first ten Principal Components of our participants' Daily Displacement Profiles. Labels on the horizontal axis are the same as Figure \ref{fig:ddp}. Cell color indicates the sign of a particular variable in relation to a PC. Red indicates that a specific PC is in the same direction as a variable whereas grey indicates that it is the opposite. Darker color (red or blue) indicates loadings with a greater absolute value.}   
    \label{fig:pca}
\end{figure}

\subsection{Principal Component Analysis} 

Once the DDPs are constructed, we treated each dimension as a separate feature or variable and conduct Principal Component Analysis (PCA) to discover the representative linear combinations of the local displacement values that can explain large proportions of the variance within the participants' movement patterns. The results indicate that 63\% of variance is explained by the first 10 Principal Components (PCs); 81\% of variance by the first 20 and 92\% by the first 30. Figure \ref{fig:pca} visualizes the weights on each displacement variable (i.e., loadings) of the first 10 PCs. The first Principal Component represents an entire day of moving around with two peaks at about 3pm and 8pm. The second PC has positive weights on displacement in the morning hours but negative weights on displacement at around 8pm. All PCs had almost zero weights on the early morning hours, during which participants were most likely sleeping. We then performed the Welch T-test on the value of each of the first 10 PCs between the pre- and post-outbreak participant groups. 

\begin{figure}[t]
    \includegraphics[width = \textwidth]{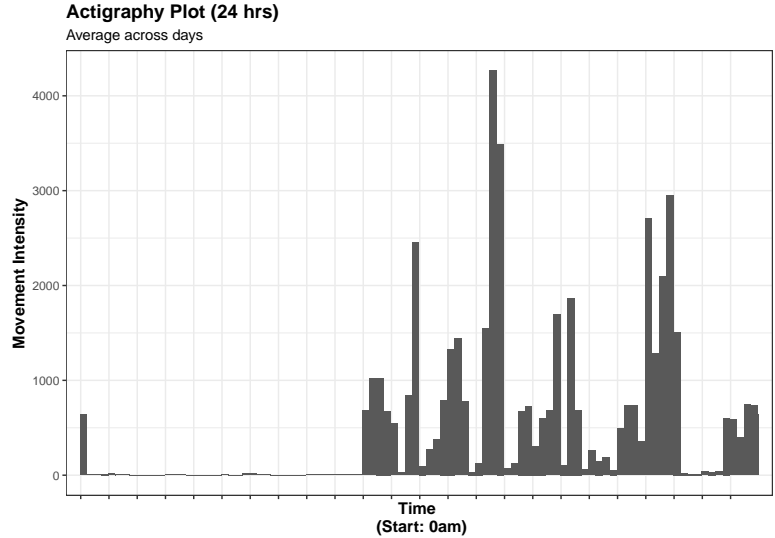}
    \caption{Distribution of half-hour displacements of one participant averaged across all days.}   
    \label{fig:cr}
\end{figure}

\subsection{Circadian Rhythm Metrics}\label{subsec:cr}

Besides relative movement across different hours of the day, we were also interested in extracting circadian rhythm metrics from the participants' GPS data. Circadian rhythm quantifies the day-to-day regularity in the magnitude fluctuation of an individual's daily routine activity (Figure \ref{fig:cr} provides an illustration) and is found to be correlated with many health and well-being statuses. We borrowed several important circadian rhythm metrics from the actigraphy analysis literature \cite{blume2016nparact} and applied them on our GPS data. Specifically, we extracted the following five metrics: 

\begin{itemize}
    \item Interdaily Stability (IS): IS quantifies the stability of rest-activity rhythms or the invariability of the rhythm between different days
    \item Intradaily Variability (IV): In contrast to IS, IV quantifies the fragmentation of a rest-activity pattern
    \item M10: the average magnitude of the signals over the ten consecutive hours that have the maximum signal magnitude
    \item L5: the average magnitude of the signals over the five consecutive hours that have the minimum signal magnitude
    \item Relative Amplitude (RA): defined by $(M10-L5) / (M10 + L5)$, reflecting the contrast between the active and inactive periods of the days.
\end{itemize}    

These metrics are again daily metrics. Similar to our analysis on PCs, we performed the Welch T-test on the value of each of the five circadian rhythm metrics between the pre- and post-outbreak participant groups.

\subsection{Digital Phenotyping}\label{subsec:dp}

The previous two methods were based on Daily Displacement Profiles. Additionally, we explored the feature engineering approach and extracted seven daily-level digital phenotypes from our participants' GPS data:

\begin{itemize}
    \item Location variance (loc.var): square root of the variance of GPS coordinates
    \item Number of significant places (num.pls): determined by an established temporal clustering algorithm \cite{kang2005extracting}
    \item Normalized entropy of time spent at significant places (ent.pls): greater values indicate more equally distributed time spent at different places, whereas lower values indicate most time spent at a small number of places
    \item Percentage of time spent at home (perc.home): home defined by the place a participant spent the most time between 12-6am during the study period
    \item Total distance traveled (total.dist): the sum of the distance between every pair of consecutive GPS coordinates registered
    \item Maximum distance covered (max.dist): the greatest distance between any two GPS coordinates registered 
    \item Routine index (routine.idx): quantifying the degree to which a participant's mobility pattern over a period of time is similar to that of the same period of time on all days, as formulated by Canzian \& Musolesi \cite{canzian2015trajectories}.
\end{itemize}

With these daily GPS features computed for all participants in both groups, we further carried out two analyses. First, like the Principal Components and Circadian Rhythm metrics, we compared the values of these daily between the pre- and post-outbreak group using the Welch T-test. Second, we built supervised learning models using these GPS features to predict the experience of severe sadness during the concurrent day. Daily sadness experience was solicited from participants in both groups via smartphone delivered surveys in the study and we consider an observation as \textit{severe sadness} when the self-reported sadness level is greater than ``quite a bit" sad. While correlating personal sensing data (especially mobility patterns characterized by GPS traces) with mood experience or mental health symptoms is a typical practice in digital health phenotyping, our objective here is to discover whether there exists a significant performance discrepancy between the two groups, since the post-outbreak participants were under influence by the pandemic and subject to systematically altered mobility patterns. We limited the data used for the predictive modeling of severe sadness to those participants who reported being ``quite a bit" sad at least twice over their study period (pre or post) because there is little point in predicting severe sadness for those who are never affected by it. 35 out of the 74 participants in the pre-outbreak group, and 27 out the 52 participants in the post-outbreak group were thus retained for the analysis. We used two machine learning methods to evaluate performance: (1) mixed-effect logistic regression with random participant effect and (2) random forest with leave-one-out cross validation per participant. Logistic regression is a typical approach for two-class classification problems and random forest has proved to be an high-performing machine learner in digital phenotyping studies.

\section{Results}

\begin{figure}[t]
    \includegraphics[width = \columnwidth]{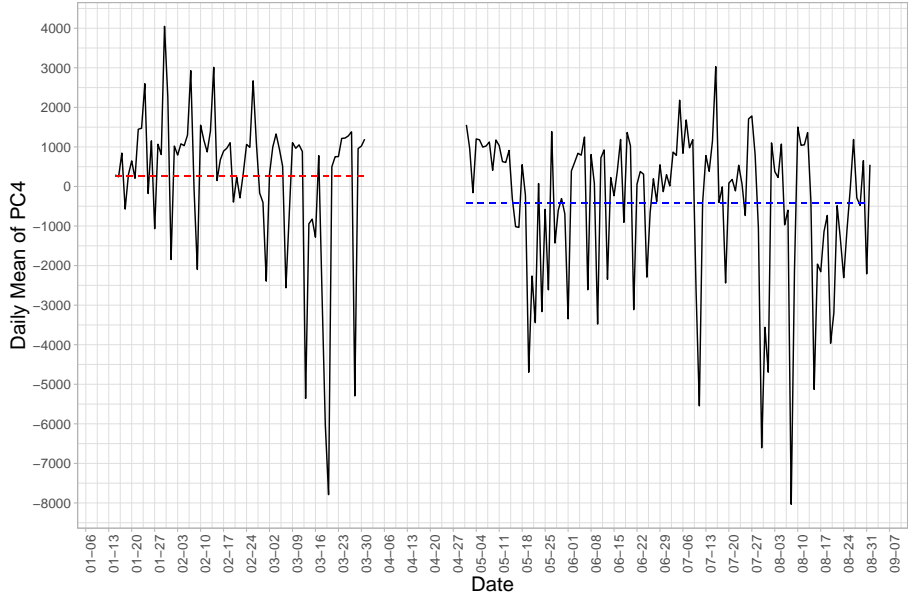}
    \caption{Daily mean value of Principal Component No.4 (PC4) over the two study periods. The red and blue dashed lines indicate the average value from the pre- and post-outbreak group respectively. } 
    \label{fig:pc4}
\end{figure}

\subsection{Principal Component Analysis}

Out of the first ten PCs extracted from our participants' Daily Displacement Profiles, one PC -- the fourth Principal Component (PC4) -- turns out to be significantly different between the pre- and post-outbreak group. PC4 is significantly lower ($p < 0.001$; $-322.4$ post vs. 406.3 pre) in the post-outbreak group than the pre-outbreak group. The remaining nine PCs do not achieve statistical significance. We aggregated the value of PC4 by its mean value by the day to show its variation throughout the two study periods (Figure \ref{fig:pc4}). The lower value of PC4 in the post-outbreak group is visible with more days below zero than the pre-outbreak group. 

As shown in Figure \ref{fig:pca}, what PC4 represents is a day with increased movement in the morning (\raisebox{-0.8ex}{\textasciitilde}8am) and in the evening (\raisebox{-0.8ex}{\textasciitilde}8pm) but decreased movement in-between. As such, participants in the post-outbreak group spent more of their days going places during the day whereas participants in the pre-outbreak group had a more concentrated mobility pattern with greater movement in the morning and in the evening.

\subsection{Circadian Rhythm Metrics}

Out of the five Circadian Rhythm metrics described in Section \ref{subsec:cr}, Intradaily Variability (IV) and Relative Amplitude (RA) showed statistical significance between the two groups whereas the remaining three did not. The post-outbreak group is significantly lower in both IV ($p=0.004$) and RA ($p=0.002$) than the pre-outbreak group. Of note, Interdaily Stability (IS), characterizing the variation in circadian rhythm over multiple days, is not significantly different between the two groups. 

Straightforwardly, a lower IV indicates less presence of rest-activity alternation within a day, and a lower RA indicates a smaller contrast between the magnitude of movement between active and inactive periods within a day. Significant lower values in both IV and RA suggest that participants experienced less ``chaotic" days in terms of mobility in the post-outbreak group.

\begin{figure}[t]
    \includegraphics[width = \columnwidth]{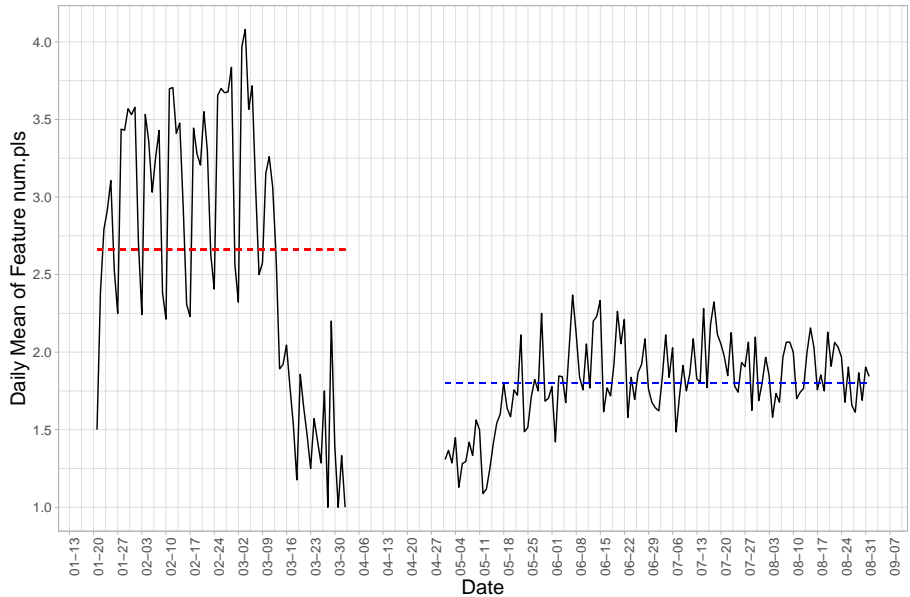}
    \caption{Daily mean value of the \textit{num.pls}, number of significant places visited, over the two study periods. The red and blue dashed lines indicate the average value from the pre- and post-outbreak group respectively. Note the drastic drop in March in the pre-outbreak group, coinciding with the declaration of pandemic in the US on March 13, 2020.} 
    \label{fig:num_pls}
\end{figure}

\subsection{Digital Phenotyping}

Out of the seven GPS features described in Section \ref{subsec:dp}, three features were significantly different between the two participant groups, namely \textit{num.pls} (number of significant places), \textit{ent.pls} (normalized entropy of time spent at significant places), and \textit{perc.home} (percentage of time spent at home). Of the three, \textit{num.pls} and \textit{ent.pls} were significantly lower in the post-outbreak group with p-values both lower than 0.001. Participants in the post-outbreak group on average visited one less significant place during their days compared to their pre-outbreak counterparts (1.73 post vs. 2.73 pre, illustrated on a day-to-day basis in Figure \ref{fig:num_pls}; note the drastic drop in March in the pre-outbreak group, coinciding with the declaration of pandemic in the president of the United States on March 13, 2020 \cite{trump2020proclamation}.). The percentage of time spent at home, on the other hand, is significantly higher in the post-outbreak group than the pre-outbreak group (20.8\% post compared to 16.3\% pre). These contrasts suggest that in the post-outbreak group participants were more home-bound and ventured outside less than before the pandemic.

As for performance of the predictive modeling tasks targeting severe sadness experience, the area under ROC value for the pre-outbreak group was 0.71 with a standard deviation of 0.13 whereas it was 0.68 for the post-outbreak group with the same standard deviation of 0.13. The direction of this difference is expected because we hypothesized that due to systematically altered mobility pattern in the post-outbreak group, the pre-established (by studies conducted during normal times) correlation between mobile sensed mobility pattern and mental health symptoms should be lower. However, the difference is not statistically significant in our experiment ($p=0.38$), possibly due to the relatively small number of participants tested upon. 

\section{Discussion}\label{sec:discussion}

Some of our findings such as the significantly increased at-home time in the post-outbreak group is consistent with findings in similar studies with entirely different participant cohorts \cite{sun2020using}. Compared to current literature, our findings offer new insights into the intradaily mobility patterns of college students during the pandemic, such as the reduced number of significant places visited during a day, the reduced Relative Amplitude in daily displacement profile from our circadian rhythm analysis, and the shifted temporal distribution of daily movement over different hours as revealed by our Principal Component analysis.

One limitation of our study is the potential confounding factor that is the eventual overlap between summer time and the timeline of the post-outbreak study period. Based on our past experience observing college students' behavior, with generally fewer constraints of academic activities, college students tend to be more active during summer time, therefore we believe our findings regarding reduced movement in our post-outbreak group still serve to accentuate the effect of the pandemic on student mobility patterns.

\section{Conclusion}

In this paper we presented our findings from a two-period smartphone sensing study we conducted using college student participants at a major US public university before and during the COVID-19 pandemic. We focused on the mobility patterns revealed by the GPS data collected from the students' smartphones and applied three analytical methods, namely principal component analysis, circadian rhythm analysis, and digital phenotyping, to characterize the differences in intradaily movement patterns between the two groups. Our findings suggest that compared to the pre-COVID group, students in the mid-COVID group (1) registered significantly more movement \textit{during} the day rather than in the morning (8-10am) and in the evening (7-9pm); (2) exhibited significantly less intradaily variability in their daily movement, and (3) had a significant lower correlation between their mobility patterns and their depression symptoms. These findings together portray a less active, less structured, and more home-bound daily movement routine of college students in the post-outbreak group and deepen our understanding of the ways college students' daily lives have been affected by the COVID-19 pandemic.

\section*{Funding}

This work was supported by Whole Communities—Whole Health, a research grand challenge at the University of Texas at Austin.

\bibliographystyle{plain}
\bibliography{main}

\end{document}